\documentclass[12pt]{article}
\usepackage[margin=0.7in]{geometry}
\usepackage{float}
\usepackage{amsmath}
\usepackage{amssymb}
\usepackage{graphicx}
\usepackage{cite}
\usepackage{color}
\usepackage{dcolumn}
\usepackage{bm}
\usepackage{newfloat}
\DeclareFloatingEnvironment[name={SI Figure}]{suppfigure}
\DeclareFloatingEnvironment[name={SI equation}]{suppequation}

\begin{document}

\begin{flushleft}
{\large
\textbf{Randomness and preserved patterns in cancer network}
}
\\
Aparna Rai$^1$,
A. Vipin Menon$^2$ and 
Sarika Jalan$^{1,2,\ast}$\\
1. Centre for Bio-Science and Bio-Medical Engineering, Indian Institute of Technology Indore, M-Block, IET-DAVV Campus,
Khandwa Road, Indore 452017, India\\
2. Complex Systems Lab, Discipline of Physics,  Indian Institute of Technology Indore, M-Block, IET-DAVV Campus,
Khandwa Road, Indore 452017, India\\

$\ast$ E-mail: sarika@iiti.ac.in
\end{flushleft}

\begin{abstract}
Breast cancer has been reported to account for the maximum cases among all female cancers till 
date. In order to gain a deeper insight into the complexities of the disease, 
we analyze the breast cancer network and its normal counterpart at the 
proteomic level.
While the short range correlations in the eigenvalues exhibiting universality provide an evidence towards the importance of random connections in the underlying networks, the long range correlations along with the localization properties reveal insightful structural patterns involving functionally important proteins. 
The analysis provides a benchmark for designing drugs which can target a subgraph instead of individual proteins.
\end{abstract}

One of the major goals of the post-genomic era is to understand the role of 
proteomics and genomics in human health and diseases \cite{Venter}. 
The world health organization has estimated that about 11 percent of the total
cancers accounts for breast cancer and a drop of about one-third of the cancer deaths could be attained if 
detected and treated early \cite{Parkin}.
Molecular studies and healthcare research has shown that the detection of more and more genes after BRCA1 and BRCA2 implicative in breast cancer has rendered research, diagnosis and treatment strategies more ambiguous as the dangers being posed by those genes are yet uncertain \cite{ScienceMarch2014}. 
Complexity as well as variations at every stage of the cancer render designing drug targets very difficult \cite{Hanahan,Petricoin}.
The ample availability 
of data in functional genomic and proteomic information and the development 
of high-throughput data-collection techniques has resulted from basic gene-based traditional 
molecular biology to a systems approach of network biology \cite{barabasi, Hartwell}. 
In this approach, biological processes are considered as complex networks of interactions between 
numerous components of the cell rather than as independent interactions involving only a few 
molecules \cite{Kitano,Zhu,Barabasi_correlation}.
Previous attempts to understand various diseases under network biology approach reveals that various types of cancers are interlinked to 
each other through some pathways which are altered in different diseases \cite{Goh}.
Further analysis of the centrosome dysfunction under the network theory framework demonstrates importance of hub proteins as well
as those connected with them \cite{Pujana,Chuang}. 
This paper, in order to achieve a deeper understanding of the complexity of breast cancer, its interacting patterns, 
role and importance of interaction patterns for the disease, analyzes protein protein interaction (PPI) 
network using a novel mathematical tool random matrix theory (RMT).
This technique is known to develop fifty years back to explain
interactions of complex nucleons \cite{Wigner},
has recently exhibited its remarkable 
success in understanding complex systems arising from diverse fields ranging from
quantum chaos to galaxy \cite{rev_rmt}. 
While structural parameters namely degree, clustering coefficient (CC), 
degree clustering correlation and diameter demonstrate similarities
in both the normal and disease networks, the structural patterns such
as cliques combined with proteins revealed through the spectral analysis
indicate changes in both the networks, which might be
important behind transformation of a cell
from the normal to the disease state.
The present work not only straightens the importance of structural patterns in the disease, 
further demonstrating the success of the network framework, but also for the first time analyzes a 
disease using the RMT techniques. This combined framework helps in detecting
proteins, beyond their structural significance in the underlying network, which are crucial for the disease.
A detailed analysis of the top contributing nodes (TCNs) in the localized eigenvectors
reveal their importance in the occurrence of the
disease state. 

\section*{Results and discussions}
\subsection*{Structural properties of cancer networks} 
The structural parameters of the largest connected cluster of the network using dataset described in the materials and methods section, are summarized in the Table.~\ref{network_parameter}.
\begin{table}[ht]
\begin{center}
\begin{tabular}{|c|c|c|c|c|c|c|c|c|}    \hline
Network 	& $N$ 	& $N_{LCC}$ & $NC_{LCC}$ 	& $\langle k \rangle$ & $D$  & $\langle CC \rangle$   & $N_{CC}$  &  $\langle IPR \rangle$  \\ \hline
Normal		& 2464	& 2441	& 15118	&  12	& $11$ & 0.28	&  153	& 0.004	\\ \hline
Disease		& 2096 	& 2046	& 14150	&  14	& $10$ & 0.29	&  107	& 0.005	\\ \hline
\end{tabular}
\caption{{\bf Network parameters for both the normal and disease networks.}The total number of proteins  $N$ collected using
various databases (described in the Method section), number of proteins
in the largest connected cluster $N_{LCC}$ and connections $NC_{LCC}$, 
the average degree $\langle k \rangle$, average clustering coefficient $\langle CC \rangle$,
 the number of nodes having $CC=1$ in the whole network ($N_{CC}$) and the average IPR of both the networks.}
\label{network_parameter}
\end{center}
\end{table}
The degree distribution $\rho(k)$ of 
both the networks follow the power law (Figure-1)
indicating the presence of very high degree nodes. 
These nodes are known to keep
the network robust against random external perturbations,
as well as have been found to be functionally important in many 
pathways \cite{PNAS2003}. The degree and CC of the normal and the disease networks
are negatively correlated (Figure-1), as found in the case of other 
biological systems \cite{Zimmer}. 
As mentioned in the Table.~\ref{network_parameter}, the 
average CC of both the networks is high as exhibited by most of the biological systems 
investigated under network theory framework indicating 
presence of functional modules \cite{Barabasi_correlation}. 
What follows that the disease and normal networks exhibit overall similar statistics
for widely investigated structural properties but the crucial differences between them which
are of potential importance, are revealed through the analysis of cliques structures and 
spectra.
As reflected in Table.~\ref{network_parameter}, the disease network bears 
less number of nodes with $CC=1$ than the normal one. 
The value of $CC$ for a node
being one reflects the formation of complete subgraph or clique comprising of that node. 
The higher value of average CC implicates the presence of high number of clique structure in a network \cite{Watts}.
Further, cliques are known as building blocks of a network for making
the underlying system more robust \cite{Alon} and stable \cite{skd}.
Additionally nodes forming cliques structure
are known to be preserved during evolution \cite{clique_mutation}.
What follows that the disease network having less cliques of order three as well as less
number of nodes with CC being one as compared to the normal one indicate that
there is a demolition of building blocks in the disease state, which may be 
leading to a more {\it unstable} underlying system, and
might be one of the reasons behind occurrence of the disease.

The importance of clique structures would become more clear after we explain
spectral properties and local patterns of top contributing nodes (TCNs)
appearing in the spectral analysis performed under the powerful
framework of RMT. This analysis not only reveals functionally
important proteins but also helps in uncovering importance of
structural pattern in the disease network.
In the following, we provide the results pertaining to the global spectral properties, eigenvalue fluctuations and properties of nodes appearing in the localized eigenvectors. These are the most frequently used techniques in RMT for analysis of spectral properties in order to achieve a comprehensive understanding of the underlying complex system.
\begin{figure}[ht]
\centerline{\includegraphics[width=0.8\columnwidth]{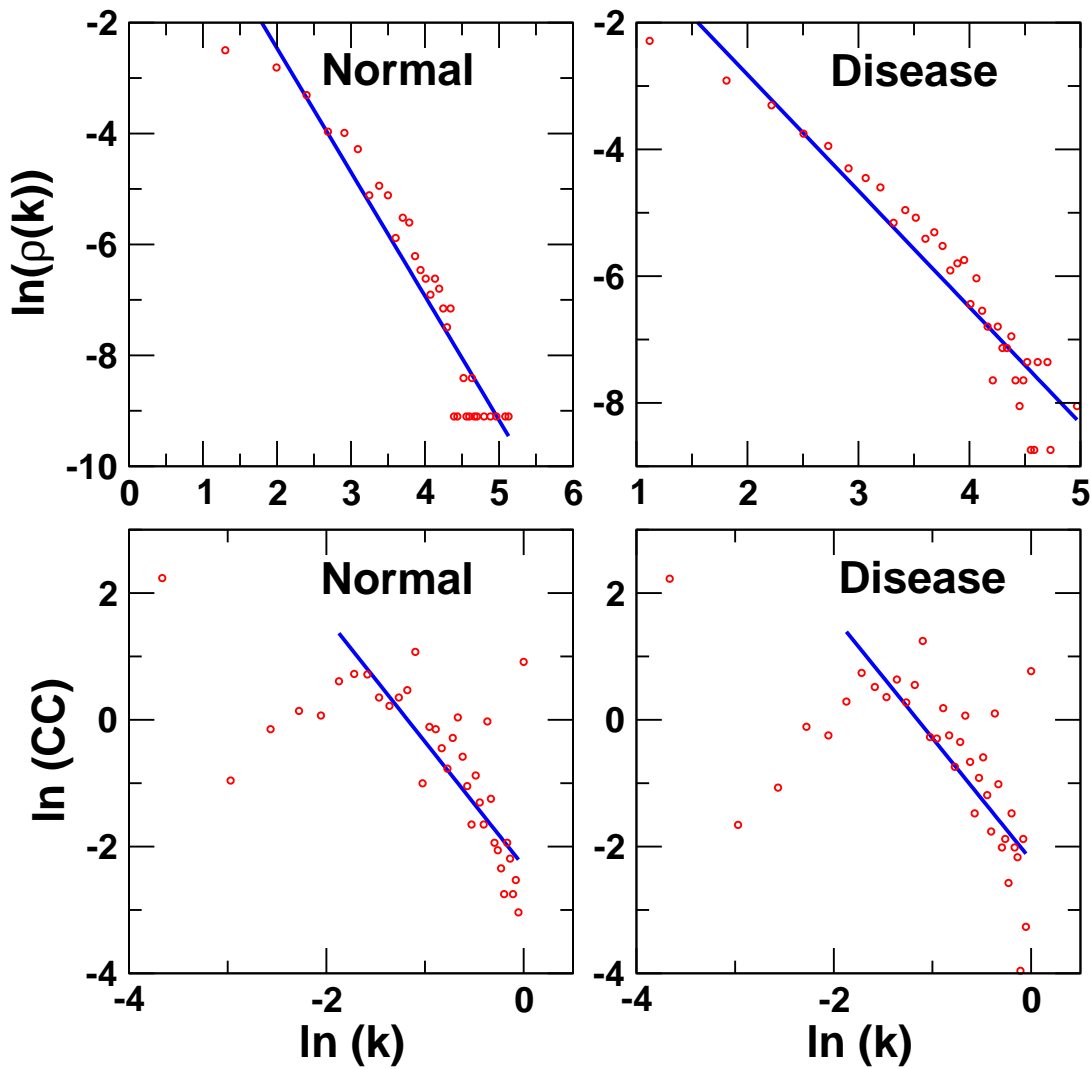}}
\caption{{\bf Degree distribution and degree-CC correlations for the normal and disease networks.} Left panel of the normal network show that the degree distribution follows power law and the degree clustering coefficient correlation shows are negatively correlated. The right panel gives us the same results for the disease networks.}
\label{structural}
\end{figure}

\subsection*{Universality and the deviation from the same}
The eigenvalue statistics reflects typical triangular shape 
with the tail of the distribution (Figure-2) relating with the exponent of the power 
law of degree distribution as observed for many other biological and real world 
networks \cite{Aguiar,SJ1}.  
Both the disease and normal networks have about $30\%$ eigenstates with 
zero eigenvalues. This high degeneracy at zero is not surprising as many of the biological networks have been shown to yield very high
degeneracy at zero \cite{SJ1}.
\begin{figure}[ht]
\centerline{\includegraphics[width=0.8\columnwidth]{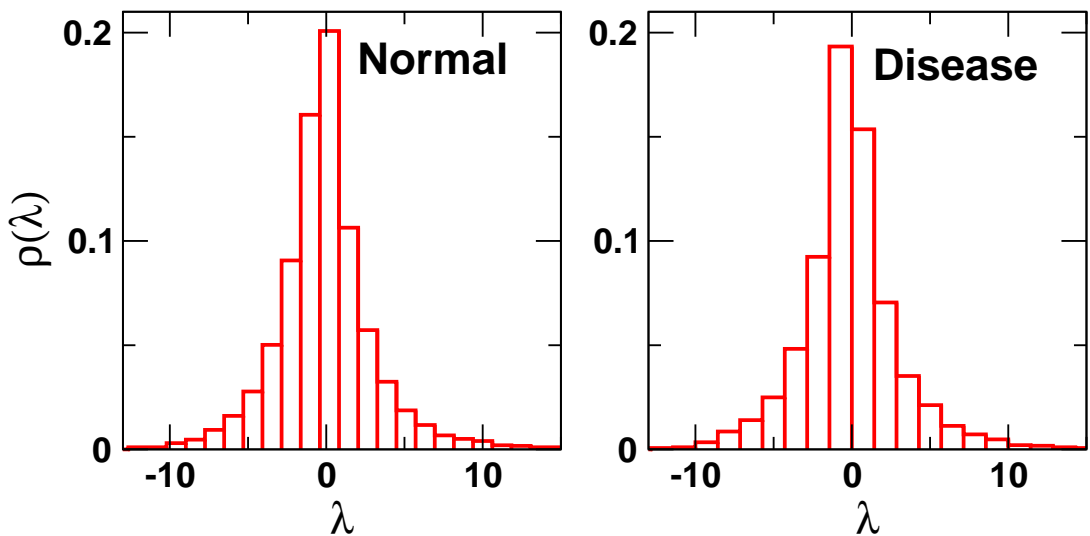}}
\caption{{\bf Eigenvalue distribution of both normal and disease networks.} The plots depict triangular distribution for both the networks with a high degeneracy at zero.}
\label{density}
\end{figure}
Further, as depicted in Figure-3, both the disease and normal 
networks follow GOE statistics of
random matrix theory at the consecutive eigenvalues captured through the distribution of their ratio (Eq.~\ref{eq_PWR}),
which reflects that both the networks have a {\it minimal amount of randomness} \cite{SJ1}. 
Randomness in a network might be arising due to some nonsense mutations 
\cite{point_mutation} occurring in the underlying system. 
We remark that in dynamical systems, randomness may be related with the unpredictable nature of 
time evolution (for example: chaotic systems) \cite{SJ_complexity}, whereas
for networks, randomness is referred to as random connections between nodes \cite{SJ_epl2009},
which for biological systems might have evinced in the course of evolution
randomly and not because of any particular functional importance of that connection. 
For instance, emergence of the modular structure in networks, which are known to be 
motivated by their specific functional role in the evolution \cite{mod_evolution}
might be linked with random connections perhaps resulting from mutations \cite{point_mutation}. 
Further, randomness in the interactions has been known to be
important for functioning of the underlying system. For instance,
information processing in the brain is considered to arise because
of many random long-range connections among different modules
\cite{Cohen} making the underlying system robust \cite{Alon}. 
(Supplementary material contains details about RMT technique).
The universal Gaussian orthogonal ensemble (GOE) statistics displayed by the disease network on one hand, 
indicates the {\it robustness} of the cell even in the disease 
state, which may be considered crucial for maintenance and housekeeping processes of cancerous cell and
on other hand establishes that the breast cell can be modeled using GOE 
of RMT and we can apply all the techniques developed under the well established framework of the
RMT to understand the breast cancer. 
\begin{figure}[ht]
\centerline{\includegraphics[width=0.8\columnwidth]{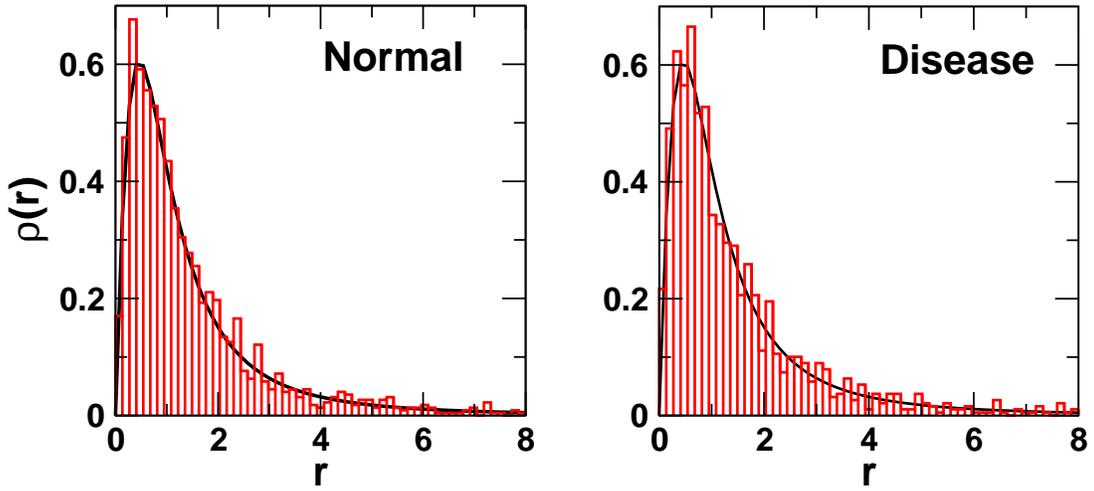}}
\caption{{\bf Spacing distribution.} The ratio of eigenvalues spacing follow
GOE statistics for both the networks. The bars
represent data points and solid line represents Eq.\ref{eq_PWR}
with parameters of GOE statistics.}
\label{ratio}
\end{figure}

After the short range correlations, which is analyzed through probability of ratio of consecutive level, 
the second most insightful step in the RMT is the analysis of
long range correlations in eigenvalue using spectral rigidity test, 
which is generally done using $\Delta_3$ statistics given by Eq.~\ref{eq_delta3}.
This test reveals that both the networks follow RMT 
prediction of GOE statistics till a particular value of $L$ (Eq.\ref{eq_delta3_goe}) and 
deviates thereafter. Interestingly, the value of $L$ for the disease network is less as compared
to the normal one (Figure-4), suggesting that the normal 
network is more random than the disease network \cite {SJ_epl2009} or  
the disease network is more ordered than
the normal one. This interpretation combined with the observation that
the disease state has a less number of connections (pathways
expressed) than that of the normal one, implicates that there are some
interactions getting hampered or silenced during the course of mutation leading
to the disease \cite{Goh}. What follows that, these hampered pathways should be corresponding to
or treated as {\it random pathways} and as {\it randomness} is one of
the essential ingredient for the robustness of a cell \cite{SJ_epl2009},
lack of {\it sufficient randomness} might be leading to the disease. 

While universal part following RMT prediction
reflects the importance of {\it random connections} in the biological networks, we will witness in the following that
the non-universal part of the spectra deviating from RMT provides
direct clue about the set of nodes (proteins) relevant for the occurrence of the disease state. This is achieved by analyzing localization properties of eigenstates which provides a quantitative picture of non universal part of the spectra. 

\begin{figure}[ht]
\centerline{\includegraphics[width=0.8\columnwidth]{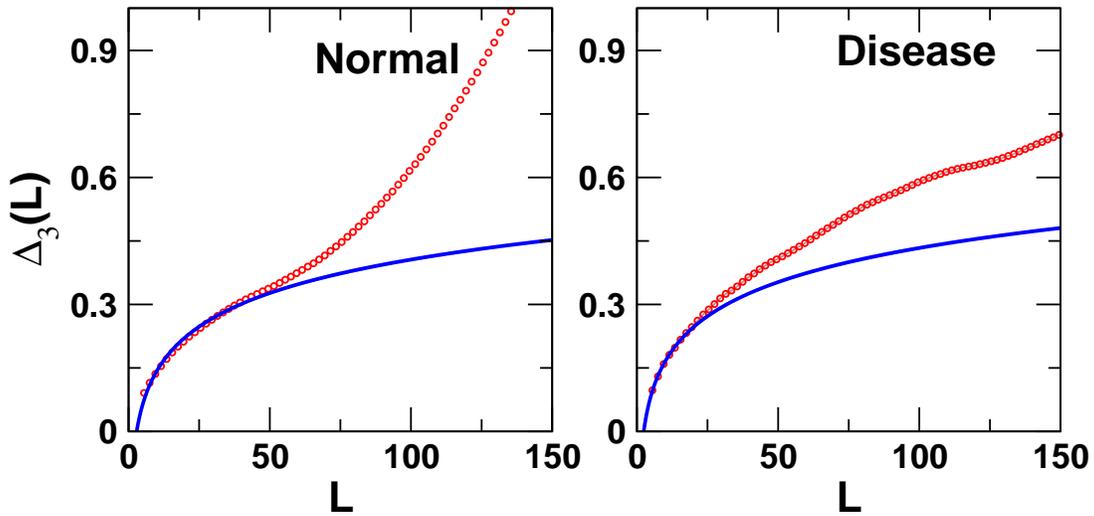}}
\caption{{\bf Long-range Correlations ($\Delta_3$ statistics) for the normal and the disease network.} Circles
denote data points for the normal and the disease networks whereas 
the solid line is the $\Delta_3$ statistics for the GOE.}
\label{Fig_Delta3}
\end{figure}
\subsection*{Important proteins through eigenvector localization}
Based on the IPR values calculated using Eq.\ref{eq_IPR}, 
the eigenstates can be divided 
into two components, one which follows RMT predictions of Porter-Thomas distribution \cite{PT},
and another one which deviates from this universality and show localization (Figure-5).
\begin{figure}[ht]
\centerline{\includegraphics[width=0.8\columnwidth]{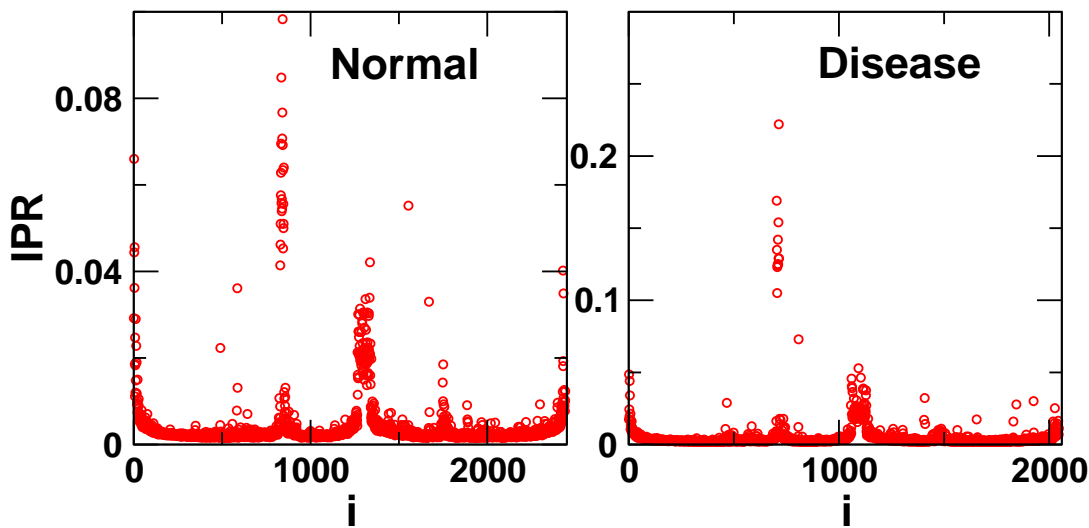}}
\caption{{\bf Eigenvector Localization both normal and disease network.} 
They clearly reflect three regions (i) degenerate part in the middle,
(ii) a large non-degenerate part which follow
GOE statistics of RMT and non-degenerate
part at both the end and near to the zero eigenvalues which deviate from RMT.}
\label{IPR}
\end{figure}

The average IPR calculated using Eq.\ref{eq_IPRavg} comes out to be more 
for the disease network than for the normal one, which is in
the direct relation with the behavior of $\Delta_3$ statistics demonstrating 
that the normal network is more random than the disease network. 
The part of the unfolded spectra following universality
corresponds to random interactions in the
underlying system \cite{SJ_gene}, whereas
non-universal part can be exploited to get the system dependent information. 
The non universal part annotating the importance of the localized eigenvectors reveals important proteins as explained below.
The disease network yields 34 TCNs corresponding to the top 5 
most localized eigenvectors extracted by taking the threshold as $1/IPR$ 
\cite{Plerou2002}, of which 18 appears to be unique in the disease and 14 are common to both the
disease and the normal networks. 
In the following section, we discuss briefly the functions of these proteins selected 
through the localization property of eigenvectors (Table.~\ref{Table-2}).

\begin{table}[ht]
\begin{center}
    \begin{tabular}{ | p{0.6cm} | p{6.0cm}| p{2cm}| p{2cm}|} \hline
$ E^k $ &  TCN &	k & CC   \\ \hline
714 &	MTHFD1L, ALDH1L1, CACNB1, CACNB2, KCNB2 &	2, 2, 3, 3, 5 & 1, 1, 0.66, 0.66, 1
	\\ \hline
704 &	KCNB2, KCNH4, TEX1, CACNB2, KIAA1, THOC1 & 	5, 5, 6, 3, 3, 3 & 1, 1, 1, 1, 1, 1
		\\ \hline
712 &	KLK5, KLK10, CNTN4, ASTN2, BRMS1L, ARID4B, KCNQ5 & 3, 3, 2, 2, 3, 3, 5 & 1, 1, 1, 1, 1, 1, 1
	\\ \hline
710 &	THOC1, BRMS1L, KCND3, KCNB2, KCNQ5, ARID4B, TEX1, KCNH4 & 6, 3, 5, 5, 5, 6, 5, 3 & 1, 1, 1, 1, 1, 1, 1, 1
	\\ \hline
705 &	KCND3, FCRL3, FCRL5, TEX1, THOC6, KCNH4, KCNQ5, MTFHD1L &  5, 2, 2, 6, 6, 5, 5, 2 & 1, 1, 1, 1, 1, 1, 1, 1
	\\ \hline
\end{tabular}
\caption{{\bf Eigenvector localization properties.}Top most localized eigenvectors ($E^k$) for the disease dataset, their top contributing
proteins and network parameters namely degree and clustering coefficient. The betweenness centrality of all the TCNs in the disease network is zero.}
\label{Table-2}
\end{center}
\vspace{-0.8cm}
\end{table}

\subsection*{Functional properties of disease proteins}
The most important outcome of the functional analysis of  
proteins corresponding to the TCNs is that all of them are involved in 
important pathways leading to
breast cancer. The first localized eigenvector has five contributing nodes, 
of which MTHFD1L is responsible for synthesis of purines in mitochondria
but its expression is up-regulated in breast cancer leading to proliferation, 
invasiveness and anti-apoptotic activity \cite{minguzzi}. The protein corresponding to the
next TCN is ALDH1L1 which in the normal cell controls the cell mobility, but in the
disease state is silenced thereby making way for uncontrolled 
proliferation of cells \cite{aldh1l1}. The next two proteins
CACNB1 and CACNB2 help in calcium transport in the normal state but are down-regulated in 
the breast cancer cell, thereby affecting the calcium metabolism which causes 
poor signaling of messages in the cell \cite{cacnb}. 
Mutation in the last TCN KCNB2, results in concomitant cell 
proliferation \cite{kcnb2}.

The second localized eigenvector consists of six TCNs, 
among which KCNB2 and CACNB2 have been discussed above. The third protein
KCNH4 in a normal cell is responsible for the potassium transport but 
undergoes splicing and is silenced in the disease state,
thereby hampering early event in apoptosis leading to prolonged survival of 
cells \cite{kcnh4}. The next protein
TEX1, under normal condition, plays a major role in transcription regulation, but in the breast cancer cells, is over expressed which causes down-regulation of transcription \cite {tex1}. 
The next TCN, KIAA1, is a binding protein to RNA but in cancer state undergoes mutation and refrains from its function of binding to RNA 
resulting in hampering of proper translation in cancer \cite{kiaa1}. The
next protein THOC1 is a part of TREX complex which is responsible for 
regulating transcription in a normal cell but in breast cancer cell, it delays transcriptional expression leading to irregular central dogma which makes the cell unorganized \cite{tex1}.

There are seven TCNs in third most localized eigenvector. The first two
proteins, KLK5 and KLK10, are from Kellikein gene family which
are generally involved in serine-type peptidase activities. In the
breast cancer they are down-regulated leading to the suppression of tumorigenesis \cite{klk}. 
The next protein CNTN4 is implicated in nervous system development. As of now its exact function in breast cancer is not known, but its first interacting neighbor is BRCA1 which is potent candidate in breast cancer cells \cite{cntn}. Similarly, the next contributing node, ASTN2, under normal 
condition controls neural migration but in the breast cancer is found to have undergone 
chromosomal rearrangement with PTPRG which is an important gene for the recognition of the cancerous state in the cell \cite{astn2}. The next protein, BRMS1L, is found to reduce expression of mRNA in 
breast cancer \cite{brms1l}. Another protein, ARID4B, functions in diverse cellular processes including proliferation, differentiation, apoptosis, oncogenesis, and cell fate determination. In the disease state, it is found to cause irregular cell formation and proliferation \cite{arid}. KCNQ5 is another TCN which is a family member of KCN which affects 
the cell proliferation of the breast cell \cite{kcnb2}.

The fourth localized eigenvector has eight TCNs of 
which KCND3 elevates the influx of potassium ions in breast cancer cells \cite {kcnd3}. 
The others appear in the three most localized eigenvector and have been already discussed.
Among the top contribution nodes
of the fifth localized eigenvector, except THOC6, FCRL3 and FCRL5, 
all other have been discussed above in different localized vector. These three 
proteins have been found in both normal and breast cell. THOC6, in normal cell accounts 
for negative regulation of apoptosis, whereas in the breast cancer cell have been found to be 
silenced \cite{tex1}. Next two proteins FCRL3 and FCRL5 appear as a part of FCRL complex 
which under normal circumstances act as an adapter of protein as well as development in immunity. 
In breast cancer, they have been found to be over expressed in immune cells 
namely WBC, thereby 
making it more robust against treatments \cite{fcrl5}.

What follows that all the proteins corresponding to the
TCNs, except few, in the five most localized 
eigenvectors 
have a major contribution in promoting breast cancer. 
Moreover, fourteen common proteins are important for the normal and the disease both but leading to 
a very different behavior of cell in the two states. Whereas
for normal cell, these common proteins are involved in major functioning of the cell 
(Supplementary), in the disease cell they all are found to be
abnormally expressed or mutated leading to the disease state.

\subsection*{Preserved structures in localized nodes}
The TCNs, in addition to the functional importance pertaining to the occurrence of
the disease state revealed, exhibits interesting structural properties. This is more remarkable
in the light that all of these TCNs
lie in the low degree regime in the networks. Moreover,
their betweenness centrality also are zero further
ruling out any trivial structural significance of these nodes. 
But importance of these nodes
based on the analysis of their interactions reveals
the existence of preserved local structural patterns. 
Most strikingly, all of them follow phenomenon of gene 
duplication \cite{Ohno}, as depicted in the Figure-6, which shows TCNs being involved in the pair formation in which first node in each pair has exactly the same
neighbors as of the second node. Most remarkably, 
there are 20 duplicates (proteins having the same number of neighbors and having more than one connection)
in the whole network of which 18 are found in the TCNs of the most localized eigenvectors. 

Further insight to this local structure is surfaced when
we analyze the interaction patterns of these proteins in 
the normal breast network. What comes out from the analysis of the 
local interaction patterns of TCNs in the normal to those of disease is that 
there is either addition(s) of new interaction in the disease state
in order to build clique of order three,
or preservation of clique structure from normal to the disease or removal of interactions 
while in the normal keeping the clique intact. 
For example, TEX1 and THOC6, retain the same clique structures in both the networks (Figure-6). There are 
other proteins, for example BRMS1L and ARID4B which shed off some connections from the
normal network (Figure-6(right)) while keeping clique structure intact in the disease state (Figure-6(left)).
Further, nodes KLK5 and KCND3 form new connection in the disease state yielding the clique structures (Figure-6(left)).
The fact that in spite of less connections of the proteins in the disease state as compared to 
the normal as well as reduction of cliques of order three from the normal to the disease, 
cliques in TCNs are preserved. This may be one of the 
reason of the poor performance of drugs targeting these proteins, as 
cliques are known to be the building blocks of a system making the underlying system robust against 
external perturbations. While the less number of nodes with $CC=1$ as well as less number
of clique of order three in the
overall disease network from the normal one implicate that there is a destruction of building block,
the conservation or addition of cliques of order three in TCNs (whose function importance for the occurrence of disease
has already been emphasized) reflects that mutation in the disease cell makes the proteins
to form a stable structure.

\begin{figure}[ht]
\centerline{\includegraphics[width=0.8\columnwidth]{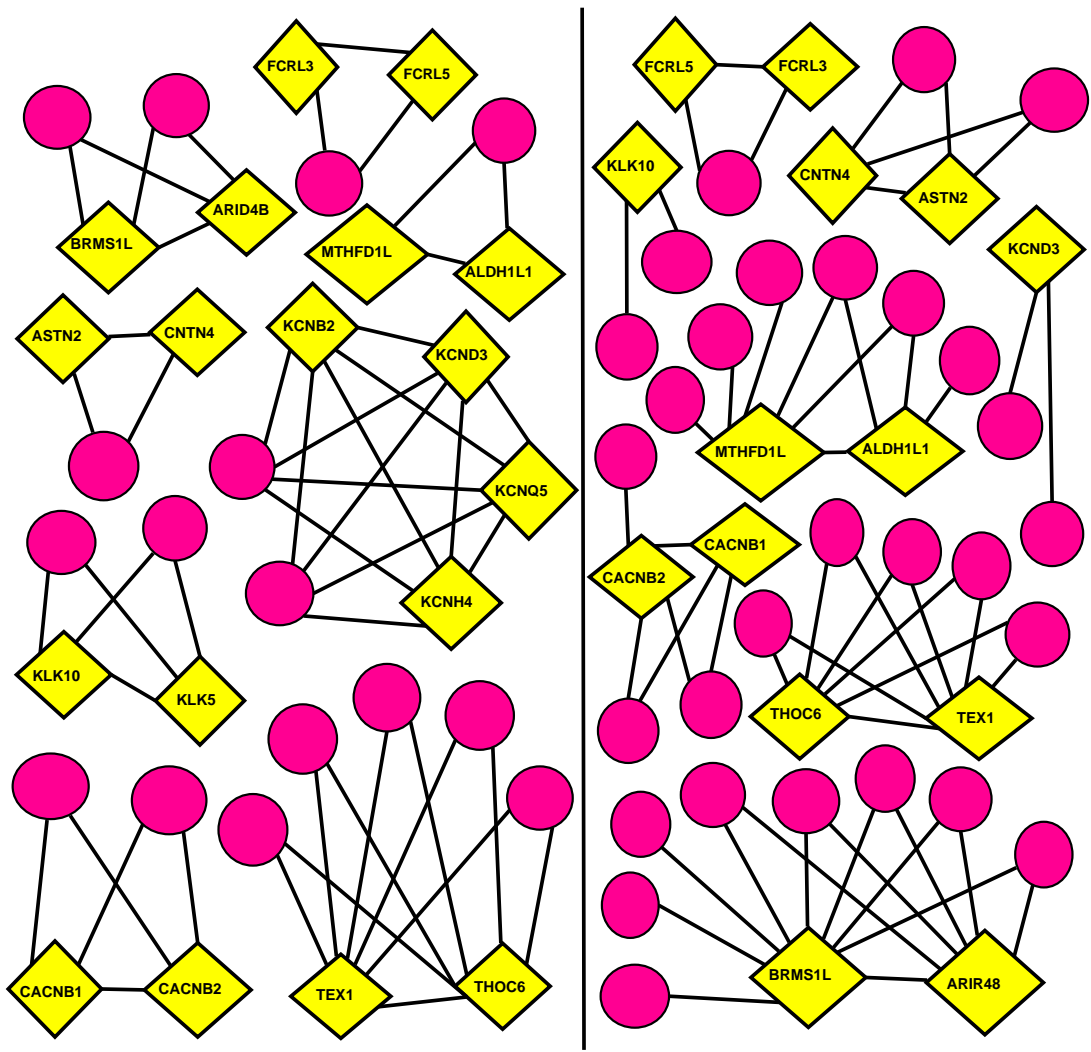}}
\caption{{\bf Local structure of top contributing nodes.} (Left panel) The local structure of all TCNs in the disease network.
(Right panel) The local structure for the same proteins in the normal network.
Yellow represents TCNs and pink represent their first neighbor.}
\label{TCN_disease}
\end{figure}

\section*{Conclusions}
We construct and investigate the breast cancer network under the RMT framework.
The analysis reveals that the TCNs of the most localized eigenvectors, despite of lying at low degree regime and having
zero betweenness centrality, exhibit structural significance.
All of them form pairs possessing common neighbors. Most remarkably, in the disease network there are 20 duplicate proteins with connections more than one, and out of which 18 appear in the localized eigenvector.  
This interesting revelation turns out to be more intriguing in the light of the
functional analysis of proteins corresponding to the TCNs which clearly confirms
their functional importance for the occurrence of the disease state.

Furthermore, clique structure and duplication of genes, which have been emphasized important for robustness as well as
the evolution of a system, are found to be crucial for the disease state.
Most striking revelation is that
while there is an overall reduction in cliques in the disease network from the normal one
reflecting reduction in the building blocks for the disease state, conservation or formation of cliques involving important disease proteins 
reflects that robustness of the overall system is decreased in the disease 
but the interactions of the important proteins involved in promoting the 
disease are preserved and might be one of the reasons behind making those pathways 
involved with the important proteins highly resistant to various treatments.

The $\Delta_3$ statistics demonstrates less {\it randomness} in the disease state than the
normal one, which might be arising due to the removal of random connections in the disease state, and
not because of a probable enhancement in the modular structure as number of nodes having
$CC=1$ is indeed less in case of the disease state than the normal one. This in-turn depicts that randomness leads to the robustness
 of the system where, normal breast network is found to be more robust than the disease state.

Detection of important proteins involved in breast cancer using RMT platform provides
a time efficient and cost effective approach for those diseases which lack in-depth information about important genes. 
Revelation of clique structure, being formed or preserved by these proteins,
provides a further bench mark for designing drugs which can target a sub-graph instead of the individual protein.  
This analysis presented here can be extended further to study other diseases like other types of cancers, diabetes, hypertension etc to predict various structural and functional aspect of biology \cite{li}, which in addition may help to compose novel drug targets and to introduce the concept of single medicine for multiple diseases \cite{ma}.

\section*{Methods}
\subsection*{Data assimilation and network construction}
According to the network theory, there are two basic components of network namely nodes and edges. 
Here we study the PPI network where nodes are the proteins and edges denote the interactions. 
After diligent and enormous efforts, we collect the protein interaction data from various literature and bioinformatic sources. 
To keep the authenticity of the data we only take the proteins into account which are reviewed and cited. 
We use various different bioinformatic databases namely Gen bank from NCBI and UNIPROT \cite{Genbank,uniprot}, 
constituting data available from other resources like European Bioinformatic Institute, 
the Swiss Institute of Bioinformatics, and the Protein Information Resource. 
To add more information we also take the most widely studied normal and breast cancer cell lines 
whose protein expression data is known. 
There are numerous cell lines available but a very few have been exploited for their maximum proteomic insight. 
Here, we are use the data of HMEC cell line for normal \cite{HMEC1,HMEC2} and MFC-7 for the breast cancer network \cite{MFC}. 
The collection and discussion to select the datasets and authenticating this data for RMT analysis is an extensive job and we generate this data after thorough literature search and validation in about 400 hrs of rigorous study. After collecting the proteins for both the datasets, their interacting partners are downloaded from STRING database \cite{STRINGdatabase}.
The dataset contains $2464$ nodes and $15131$ connections for normal network followed by
$2096$ nodes and $14183$ links for the breast cancer network. The networks turned to 
have one big cluster and several small clusters. We then investigate the structural properties of the network. 

\subsection*{Structural measures}
Several statistical measures are proposed to understand specific features of the network \cite{Barabasi_2002}. First we define the interaction matrix or the adjacency matrix of the network as 
follows:
\begin{equation}
A_{\mathrm {ij}} = \begin{cases} 1 ~~\mbox{if } i \sim j \\
0 ~~ \mbox{otherwise} \end{cases}
\label{adj_wei}
\end{equation} 
The most basic structural parameter of a network is the degree of a node ($k_i$),
which is defined as the number of neighbors of the node has ($k_i=\sum_j A_{ij}$). 
Degree distribution $\rho(k)$, revealing the fraction of vertices with the degree k, 
is known as the fingerprint of the network. 
Another important parameter is the clustering coefficient (CC) of the network. Clustering is defined as the ratio of the number of connections a particular node is having by the possible number of 
connections the particular node can have. These are also known as cliques. Clustering coefficient of a network can be written as 
\begin{equation}
CC = \frac {1} {n} \sum_{i=1}^{n} C_i
\label{eq_clique}
\end{equation}
They are complete sub graphs in the network which are known to be the conserved part of the network \cite{clique_mutation}. The average clustering coefficient of the network characterizes the overall tendency of nodes to form cluster or groups \cite {Barabasi_2002}.
Further, the betweenness centrality of a node $i$ is defined as the fraction of shortest paths between node pairs that pass through the said node of interest \cite{Newman_2003}
\begin{equation}
x_i=\sum_{st} \frac{n^i_{st}}{g_{st}},
\end{equation} 
where $n^i_{st}$ is the number of paths from $s$ to $t$ that passes through 
$i$ and $g_{st}$ is the total number of paths from $s$ to $t$ in the network.
Another parameter is the diameter of the network which measures the longest of the shortest path 
between the two nodes. 

\subsection*{Spectral techniques}
The random matrix analysis of the eigenvalue spectra considers  (1) global
properties such as spectral distribution of eigenvalues $\rho(\lambda)$,
and (2) local properties such as eigenvalue fluctuations around $\rho(\lambda)$. 
We denote the eigenvalues of a network by $\lambda_i = 1,\hdots ,N$ and 
$\lambda_1>\lambda_2> \lambda_3> \hdots > \lambda_N$. 
The nearest neighbor spacing distribution (NNSD) has been known to be one of the most powerful technique in RMT and we analyze them by calculating  
the distribution of ratio of the consecutive eigenvalues which is represented as \cite{Atas}
\begin{equation}
P_{W}(r) = \frac{1}{Z_{\beta}}\frac{(r + r^2)^\beta}{(1+r + r^2)^{(1+3/2\beta)}}
\label{eq_PWR}
\end{equation}
The benefit of analyzing ratio of nearest spacings  over much used NNSD \cite{Mehta_book} is that this method does not require unfolding
of the eigenvalues.
Further, the NNSD accounts only for the short range correlations in the eigenvalues. 
We probe for the long range correlations in eigenvalues using $\Delta_3(L)$ statistics
which measures the least-square deviation of the spectral 
staircase function representing average integrated eigenvalue density 
${N}(\bar\lambda)$ from the best fitted straight line for a finite interval of 
length $L$ of the spectrum. In order to get the universal properties of eigenvalues through
h $\Delta_3$ statistics, it is customary in RMT to unfold eigenvalues by a transformation 
$\bar{\lambda_i} = \bar{N}(\lambda_i)$, where $\bar{N}$ is the average integrated eigenvalue density. 
Since we do not have any analytical form for $N$, we numerically unfold the spectrum by polynomial curve fitting. After unfolding, average spacings are unity, independent of the system,which in the absence of any analytical form of polynomial fitting of
the eigenvalues works with the approximate numerical fitting \cite{Mehta_book}
and is given by
\begin{equation}
\label{eq_delta3}
\Delta_3 (L;x) = \frac{1}{L} min_{a,b}   \int_x^{x+L} [N(\overline{\lambda})-a\overline{\lambda}-b]^2 d\overline{\lambda}
\end{equation}
where $a$ and $b$ are regression coefficients obtained after least square fit \cite {Mehta_book}. Average over several choices
 of x gives the spectral rigidity, the $\Delta_3(L)$. In case of GOE statistics, the $\Delta_3(L)$ depends logarithmically on L, i.e.
\begin{equation}
\Delta_3(L)  \backsim \frac{1}{\pi^2} \ln L
\label{eq_delta3_goe}
\end{equation}

Further, we use the inverse participation ratio (IPR) to analyze localization properties of the eigenvectors 
\cite{Haake}. For $E_l^k$ denoting $l$th
component of $k$th eigenvector $E^k$, the IPR of an eigenvector can be defined as 
\begin{equation}
I^k = \frac{ \sum_{l=1}^{N} [E_l^k]^4}{ (\sum_{l=1}^{N} [E_l^k]^2)^2}
\label{eq_IPR}
\end{equation}
which shows two 
limiting values : (i) a vector with identical components $E_l^k \equiv 1/\sqrt{N}$ has $I^k=1/N$, whereas (ii) 
a vector, with one component $E_1^k=1$ and the remainders zero, has $I^k=1$. Thus, the IPR quantifies the 
reciprocal of the number of eigenvector components that contribute significantly. 
We further calculate the average IPR in order to measure an overall
localization of the network calculated as \cite{SJ_directed}, 
\begin{equation}
\langle IPR \rangle = \frac{ \sum_{l=1}^{N} [I^k]}{N}
\label{eq_IPRavg}
\end{equation}
Note that IPR defined as above separates out the TCNs
by keeping the threshold as $1/IPR$.

\section*{Acknowledgments}
SJ is grateful to Department of Science and Technology, Government of India
and Council of Scientific and Industrial Research, India project grants 
SR/FTP/PS-067/2011 and 25(0205)/12/EMR-II for financial support.
AVM acknowledges IIT Indore for providing a conducive environment
for carrying out his internship. It is a pleasure to thank Baowen Li for interesting 
discussion on Cancer networks and V. K. B. Kota for useful suggestions 
on eigenvalues spacings during CNSD 2013. SJ thanks Akhilesh Pandey for valuable discussions on random matrix theory.
AR is thankful to the Complex Systems Lab members Aradhana Singh, Camellia Sarkar and Sanjiv K Dwivedi for helping with data download and discussions.

\subsection*{Author contributions} 
SJ conceived and supervised the 
project. AR and AVM collected the data. SJ and AR performed the
analysis and wrote the manuscript.

\subsection*{Additional information}

{\bf Supplementary material} accompanies this paper at http://www.nature.com/scientificreports.
{\bf}\\\\
{\bf Competing financial interests:} The authors declare no competing financial interests.

\cleardoublepage
\centerline{\large \bf Supporting Information} 
\vspace{1cm}

Breast cancer being the highest among all the cancers is a potential field of research. We apply RMT to understand the biological insights of the disease through network modeling. We construct both the networks of which the unconnected cluster can be seen in the figure (SI Figure~\ref{network}).

\begin{suppfigure}[ht]
\centerline{\includegraphics[width=0.46\columnwidth]{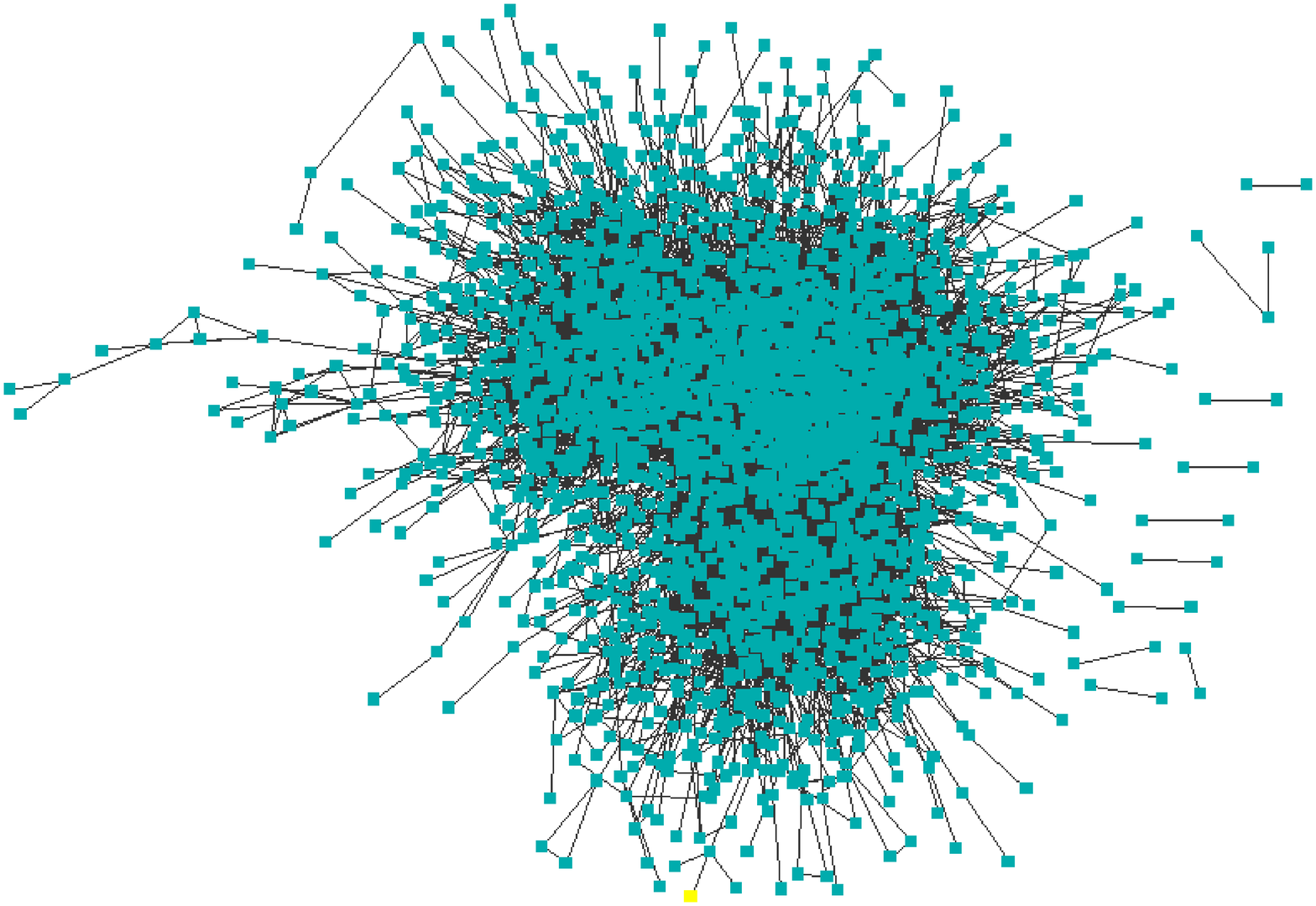}
\includegraphics[width=0.46\columnwidth, height = 5cm]{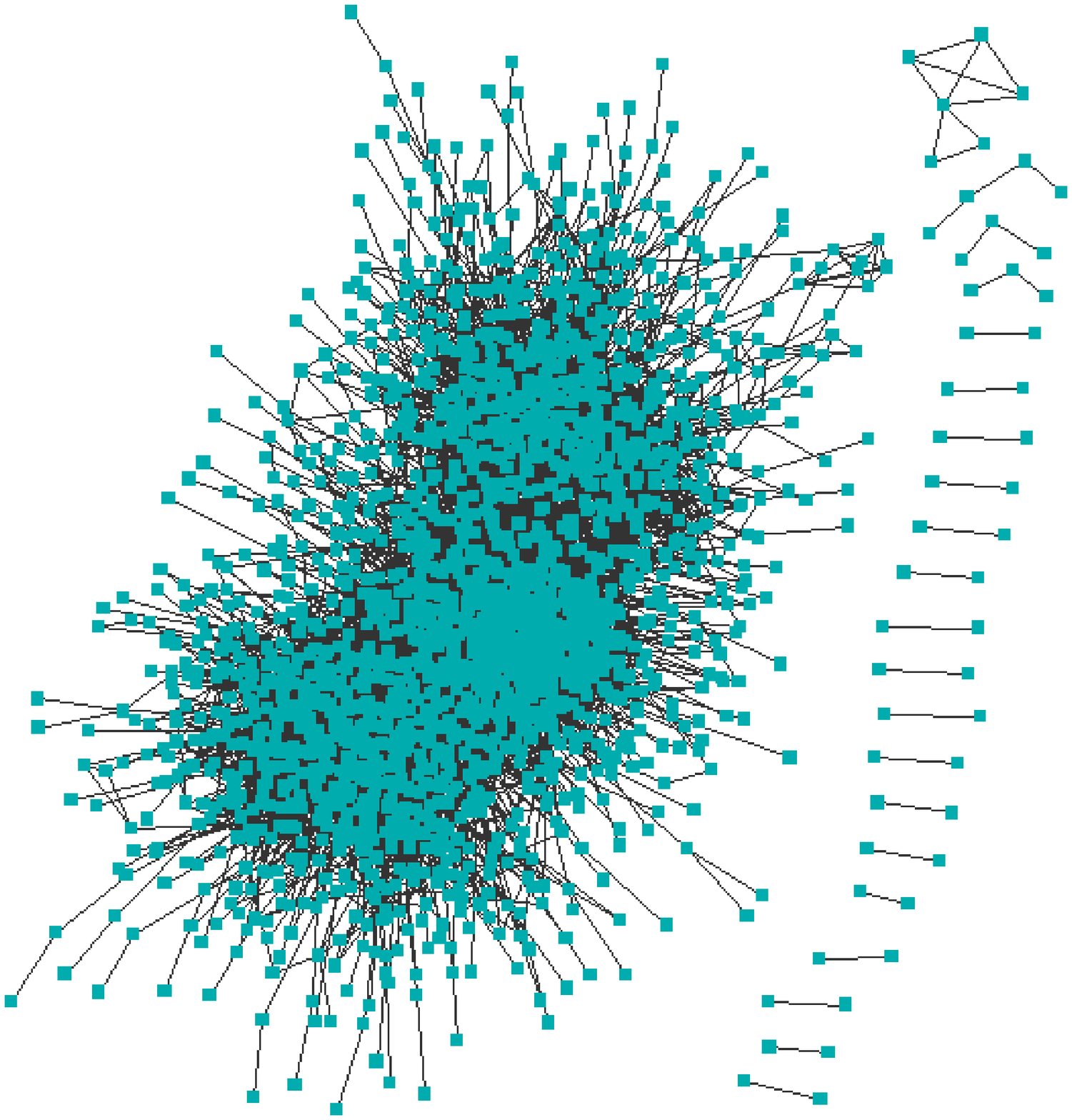}}
\caption{Normal (left panel) and disease (right panel) breast cancer network.}
\label{network}
\end{suppfigure}

\section*{Short range correlations in Eigenvalues}
 The nearest and next-nearest spacing distribution (NNSD and nNNSD) account for the short range correlations in the eigenvalues. We calculate the NNSD of normal and breast cancer network and find that it follows random matrix predictions. 

As mentioned in the main article, we denote the 
eigenvalues of a network by $\lambda_i,\,\,i=1,\dots,N$, where $N$ is size of the network and $\lambda_1 < 
\lambda_2 < \lambda_3 < \dots < \lambda_N$.  The density distribution $\rho(\lambda)$  
follows semi-circular distribution for GOE statistics. For spacing 
distribution of eigenvalues, one needs to unfold the eigenvalues by a
transformation $\overline{\lambda}_i = \overline{N} (\lambda_i)$, where
$\overline{N} (\lambda) = \int_{\lambda_{\mbox{\tiny min}}}^\lambda\,
\rho(\lambda^\prime)\, d \lambda^\prime$ is the averaged integrated eigenvalue
density \cite{Mehta}. In the absence of an analytical form for $\overline{N}$, spectrum is unfolded
numerically by polynomial curve fitting and spacings are calculated as $s_1^{(i)}= \bar{\lambda}_{i+1} - 
\bar{\lambda_{i}}$. In the case of GOE statistics, the spacing distribution is given by
$P(s_1) = (\pi/2) s_1 \exp(- \pi s_1^2/4)$
For intermediate cases, the spacing distribution is described by Brody distribution as
%\begin{suppequation}

\centerline
{$P_{\beta}(s_1)=As^\beta\exp\left(-\alpha s_1^{\beta+1}\right)$ ~~~~~~~~~~~~~~~~~~~~~~~~~~~~~~~~~~~~~~~~ (SI $1$)}
%\label{Eq_Brody}
%\end{suppequation}

where $A=(1+\beta)\alpha$ and $\alpha=[{\Gamma{({\beta+2}/{\beta+1})  }}]^{\beta+1}$.
As $\beta$ goes from 0 to 1, the
Brody distribution smoothly changes from Poisson to GOE. 
We find that the nearest neighbor
spacing distribution (NNSD) can be fitted using the Brody distribution (SI Eq.1)
%~\ref{Eq_Brody}) 
suggesting that the network can be modeled using appropriate ensemble of RMT. Fitting with the Brody distribution yields the value of $\beta=1$, for both the networks indicating that the system lies close to the GOE (Fig.~\ref{spacing}). The next nearest spacing distribution (nNNSD) $P(s_2)$, where $s_2=\bar\lambda_{i+2}-\bar\lambda_i$, 
provides a fairly good fitting with the
NNSD of Gaussian symplectic ensemble (GSE), given by

%\begin{suppequation}
\centerline
{$P(s_2)=\frac{2^{18}}{3^6\pi^3}s_2^4\exp(-\frac{64}{9\pi}s_2^2))$ ~~~~~~~~~~~~~~~~~~~~~~~~~~~~~~~~~~~~~~~~ (SI $2$)}
%\label{Eq_GSE}
%\end{suppequation}

confirming further that the underlying systems can be modeled by using universal GOE class of RMT (SI Figure ~\ref{spacing}). 

\begin{suppfigure}[ht]
\centerline{\includegraphics[width=0.8\columnwidth]{FigureS3.eps}}
\caption{(Color online) (a,b) NNSD and (c,d) nNNSD. 
The histogram represents the numerical result and the 
solid line (blue) is fitted by (a,b) Brody distribution given by SI Eq.1
%~\ref{Eq_Brody} 
and (c,d) NNSD of GSE random matrices.}
\label{spacing}
\end{suppfigure}

\section*{Functional properties of normal proteins}
\noindent The top contributing node for the first localized eigen vector are as follows ATP13A1, LACTB, ASTN2, CNTN4, TPD52L2, MAGEC2, MAGEA12, C17orf49, ALDOA, GABRA5, PTP4A1  respectively which comes around eleven contributing nodes. The first top contributing node is ATP13A1, which are transporters who are used to transport inorganic cations and other substrates through cell membranes. LACTB major role is to promote intramitochondria membrane organization and micro-compartmentalization, thereby giving power to cells. Both ASNT2 and CNTN4 are described in separate section. TPD52L2 has major role is in alternate splicing and regulatory binding of proteins in cell. MAGEC2, their role is in immune system as potential target antigen. MAGEA12 which represent antigen in cytokines and T cell of lymphocytes. C17orf49 are basically open reading frame for translation process in cell. ALDOA are enzyme associated with process of glycolysis. GABRA5 are used as inhibitory neurotransmitter in Central Nervous System. PTP4A1, major role is in cell migration and growth.

The top contributing nodes  for second most localized vector are MUCL1, GALNT5, TPD52L2, FCRL5, FCLR3, DSG2, TMEM43, NEK8, PKHD1, TPD52, VAMP7 and USE1 respectively and comes around twelve contributing nodes. MUCL1 is coded for primary transcripts in normal breast tissue. GALANT5 a part of N-acetylgalactosaminyl transferase,in normal cells serve as charge density and maintenance of B-domain. TPD52L2 was present in first localized eigen vector and has been discussed. FCRL3 and FCRL5 have been found in both disease and normal cells, hence they have been discussed in different section.DSG2 helps in epithelial morphogenesis and cell positioning in normal breast cell. TMEM43 act as enzyme protein in Wnt/Beta-catenin signal transduction pathway. NEK8 plays a major role in G2/M progression and promote centrosome maturation during mitosis. PKHD1 codes for multiple transcription protein domains. TPD52, they are part of D52 family whose primer function has been the regulation of vesicle trafficking and exocytotic secretion. VAMP7 function is lysosomal exocytosis in normal breast cell. USE1, function in the fusion of retrograde transport vesicles.

The top contributing nodes for third most localized vector are JARID2, EHMT1, SLAMF8, SLC7A7, SPG21,
ZFYVE26, C17orf49, INO80C, MGST3, MGST1, TEX1 and THOC6 respectively. JARID2 are responsible for transcriptional repression in normal breast cell. EHMT1, causes methylation of euchromatin in normal cells. SLAMF8, responsible for inflammatory reaction in normal cells. SLC7A7, primer role is metabolism of protein, especially in arginine metabolic pathway. SPG21 current role is unknown but their interacting partner ALDH1, which is factor for cell growth, cell differentiation and cell proliferation. ZFYVE26, are phosphate binding protein who are used to check cytokinesis in cells.C17orf49, has been discussed in first localized eigen vector. INO80C, helps in remodelling transcription,replication and repairment.MGST1 and MGST3, are used to control mRNA expression in normal cell.TEX1 and THOC6 are have been discussed in seperate section,since they are present in both disease and normal cell.

The top contributing nodes for fourth most localized vector are 
JARID2, EHMT1, THOC6, TEX1, ASTN2, CNTN4, TPD52L2, TPD52, SPG21, ZFYVE26, NEK8, PKHD1, SLC7A7, SLAMF8, MGST3 respectively. JARID2 , EHMT1, SPG21, ZFYVE26, SLC7A7, SLAMF8, MGST3 have been discussed in third most localized vector. THOC6, TEX1, ASTN2 and CNTN4 have been discussed in another section. TPD52L2, TPD52, PKHD1, NEK8 have been discussed in second most localized vector.

The top contributing nodes for fifth most localized vector are as follows
MGST3, MGST1, ATP13A1, LACTB, FCRL5, FCRL3, ASTN2, CNTN4, DOCK6, DOCK9, VAMP7, USE1, TPD52L2, TPD52 and MAGEC2. FCRL5, FCRL3, ASTN2 and CNTN4 have been discussed in another section. ATP13A1, LACTB, MAGEC2  and TPD52L2 have been discussed in first localized eigen vector. VAMP7,USE1 and TPD52L2 have been discussed in second most localized eigen vector. MGST3 and MGST1 have been discussed in  third localized eigen vector.DOCK6 is resposnsible for actin cytoskeleton organization in cell.DOCK9 are GTPase activating protein.

\section*{Functional Properties of first neighbours :normal}

The first neighbours normal are found to be
GPX3, GSR, TUB, GSTP1, GSTK1, GSTZ1, GGT1, ALDH1A3 are the first neighbour protein for ATP13A1, LACTB, MGST3 and MGST1. GSR also called as human glutathione reductase has key role gluthione dependent antioxidant system ,recently it has been found that, they are invovled in variation of expression in proteins.[2] TUB are those protein whose function is considered as adaptor linking insulin receptors [3]. GSTP1  glutathione-S-Transferase whose primer function is depended on glutathione but in recent time has been found out that due to hypermethylation  in GSTP1, they cause transcription silencing [4]. The kappa class of Gluthione-S-Transferase have been associated with insluin secretion along with repression in transcription, whenever hypermethylation occurs [5]. GSTZ1 Glutathione-S-transferase Zeta class are involved in enzymatic activity of glutathione transferase, but mutation in them causes functional loss as well as silencing of transcription[6]. GGT1 is used as chemical markers for identifying tumour [7]. All glutathione are used as enzymes which actively takes  part in catalysis process, apart from that mutation in them directly effects the transcription or translation process. ALDH1A3 has major role in detoxification of aldehydes and lipids, apart from that they are used as markers in Stem cells for the cure of breast cancer [8].
THOC4, THOC5, THOC7, THOC1, THOC2 are the first neighbouring protein is TEX1 and THOC6. These are transcription complex which has a predominant role mRNA transcription profiling and  complex [9]. In normal cells their basic functioning includes mRNA export.

Next comes the proteins namely INO80C and C17orf49 which has KIAA1267, MAX, MCRS1, LAS1L and HCFC1 as first neighbours. KIAA1267 along with C17orf49 are associated with chromosome 17 which is responsible for friendly behaviour, hypotonia and intellectual disability [10]. MAX are specific proteins who are involved in recurring abnormalities in human tumour [11]. MCRS1 enable the antagonist to decrease the basal level of cAMP production, hence G receptor [12]. LAS1L is nucloer protein who is required for cell proliferation and  Ribosome Biogenesis [13]. HCFC1 is a cellular factor which interacts with each protein to assemble the stable enhancer complex, in addition to the function of house keeping genes [14]. 
PTP4A2, CDC42 and RAC1 are the first neighbouring protein for PTP4A1, DOCK6 and DOCK9. PTP4A2 are used in development of placenta and embroys[15]. CDC42 and RAC1 has major role in GTP binding of protein kinase pathway [16].
GALNT6 and MUC1 are the neighbours of FCRL3, FCRL5, MUCL1 and GALNT5. GALNT6 is involved in fibronectin pathway which stabilises the development of fibronectin in breast cells [17], whereas MUC1 when binds with cell adhesion molecules are responsible for the increase or decrease in disease progression [18].

HDAC2, EZH2 and EED are the neighbours of JARID2 and EHMT1. HDAC2 and HDAC1 are protein complex which along with E-cadherin promoter cause the repression in E-cadherin expression and transcription regulation [19]. EZH2 controls DNA methylation. EED controls morphologic movements during gastrulation in embryonic development [20].
MAGEA10, MAGEA4 and MAGEA1 are the neighbours of MAGEC2 AND MAGEA12. MAGEA10, MAGE4, MAGEA1 are member of MAGE family, which are highly tumour specific [21]. 
Next we find that proteins namely,
PKD2 and NPHP3 are the neighbours of PKHD1 and NEK2. PKD2 are polycystin, which are bunch of large glycoprotein that causes cell to cell adhesion and Plexin transcription promoter region [22]. NPHP3 are nephronophthisi group which code for nephrocystein protein [23].
DSP and PKP2are neighbours of DSG2 and TMEM43 [24]. PKP2 have been implicated in cell cycle regulation and Ras signalling through its interactions [25]. DSP are potential candidates for intracellular adhesion but due to haplo-insufficiency mutation causes cardiomyopathy.

Further, GABRB2, GABRA4 and GABRA6 are found to be the neighbours of GABRA5. They represent the family of gammaaminobutryic acid, hence they represent neurotransmitters [26]. 
SPAST is the neighbour of SPG21 and ZFYVE26. Mutation in SPAST causes autosomal herditary mutation [27].
SLCO2B1 is the neighbour of SLAMF8 and SLC7A7. Its basic function is anion tansportation of polypeptide [28].
HP and CDH10 are the first neighbour of CNTN4 and ASTN2. HP are Haptoglobin complex along with haemoglobin undergo metabolism in reticuloendothelial system [29]. CDH10 are associated with neural cell adhesion molecules, it has been found out that single mutation in mRNA expression causes autism spectrum disorder(ASD) [30].
STX5 is the first neighbour of USE1 and VAMP7. STX is used for identifying the effects of nerve permiability during influx and efflux of sodium ions in the body [31].
TPD52L1 is found to be neighbour of D52 family which are responsible for regulating protein stability and transcription regulation[32,33] 

Conclusion:
Here we are able to find that most of the proteins which are in first neighbour  to be focussed on protein complexes and cell regulation whereas their parent are more focussed on house keeping gene.In a way the technic is trying to reveal that the top contributing nodes  or their first neighbours are involved in transcription complexes and regulation of cells (Cell growth proliferation etc).

\section*{Functional Properties of first neighbours :disease}

THOC4, THOC5, THOC2, THOC1 and THOC7 are the first neighbours of TEX1 and THOC6. TEX1 and THOC are basically transcription complex,and these neighbours are found to be part of it [34]. Therefore the major function is basically negative regulation of mRNA transcription, but is silenced in breast cancer.
KCNA5 and KCNA10 are found to be the first neighbours of KCNB2, KCND3, KCNQ5 and KCNH4. The above mentioned are potassium gated ion channels, major function being efflux and influx of potassium ions.In breast cancer  the efflux of potassium is exponentially high %[2][3][4].
MUC1 has its first neighbors as FCRL3 and FCRL5 which are responsible for the development in immunity, MUC1 to is overexpressed in breast cancer as immunoreceptor antigen. They are considered as potential macrophage-restricted receptor. %[5][6][7].
GART has neighbour namely ALDH1L1 and MTHFD1L, former which cause uncontrolled proliferation in breast cancer and latter causes differences in microRNA regulation. GART are responsible for brain development, but their function in breast cancer are still not known.%[8][9][10].
ASTN2 and CNTN4 are the neighbors of CDH10 where ASNT2 undergoes chromosomal rearrangement with PTPRG, there by causing further mutation in breast cells. CNTN4 plays an important role in neural development but CDH10 as their first neighbour acts as transcription regulator in tumour or breast cancer. %[11][12][13]
AKAP6 and CACNA1A are neighbours for CACNB1 and CACNB2. AKAP6 are familial proteins who in breast cancer cause extended proliferation of cells and uncontrolled growth. CACNA1A are calcium mediated channels whose efflux decreases in breast cancer.%[14][15][16].

KLK7 and KLK8 from Kalkerin gene family, are downregulated in breast cancer. They are neighbours for both KLK5 and KLK10 and encode serine protease enzyme which are preferentially downregulated.%[17]
BRMS1L and ARID4B are the neighbours for BRMS1 and ARID4A repectively. ARID4A and ARID4B are potential leukamania Suppressor gene. ARID4A and ARID4B act as adapters to recruit mSin3A-histone deactylase to E2F dependent promoters, which are silent in breast cancer. Both BRMS1L and BRMS1 are considered for the supression of metastasis. %[18][19].

\end{document}